\documentclass[english,aps,pra,reprint,noshowpacs,superscriptaddress]{revtex4-1}   
\usepackage[T1]{fontenc}	
\usepackage[latin9]{inputenc}	
\usepackage{geometry}		
\usepackage{graphicx}
\usepackage[above,below]{placeins}	
\usepackage{times}
\usepackage{hyperref}  
\hypersetup{colorlinks=true,urlcolor=blue,citecolor=blue,linkcolor=blue}   
\usepackage{color}

\begin{document}

\title{Using  Social Network Analysis on classroom video data}

\author{Katarzyna E. Pomian}
\affiliation{Department of Physics, DePaul University, 2219 North Kenmore Avenue Suite 211, Chicago, IL 60614, USA}

\author{Justyna P. Zwolak}
\thanks{{\it JPZ currently at QuICS, UMD, College Park, MD 20742, USA and NIST, Gaithersburg, MD 20899, USA}}
\affiliation{Department of Teaching and Learning, Florida International University, 11200 SW 8th Street, Miami, Florida 33199, USA}
\affiliation{STEM Transformation Institute, Florida International University, 11200 SW 8th Street, Miami, Florida 33199, USA}

\author{Eleanor C. Sayre}
\affiliation{Department of Physics, Kansas State University, 1228 N. 17th St., Manhattan, KS 66506, USA}

\author{Scott V. Franklin}
\affiliation{Department of Physics, Rochester Institute of Technology, 1 Lomb Memorial Dr, Rochester, NY 14623, USA}

\author{Mary Bridget Kustusch}
\affiliation{Department of Physics, DePaul University, 2219 North Kenmore Avenue Suite 211, Chicago, IL 60614, USA}


\begin{abstract}
We propose a novel application of Social Network Analysis (SNA) using classroom video data as a means of quantitatively and visually exploring the collaborations between students.
The context for our study was a summer program that works with first generation students and deaf/hard-of-hearing students to engage in authentic science practice and develop a supportive community. 
We applied SNA to data from one activity during the two-week program to test our approach and as a means to begin to assess whether the goals of the program are being met.
We used SNA to identify groups that were interacting in unexpected ways and then to highlight how individuals were contributing to the overall group behavior. We plan to expand our new use of SNA to video data on a larger scale.
\end{abstract}

\maketitle

\section{Introduction}\label{sec:intro}
Many studies have shown that student persistence and retention in college is strongly related to students' sense of belonging and community~\cite{Tinto1975,Tinto1997}. 
In recent years, there have been several programs developed to engage students in authentic science practice while fostering supportive communities \cite{Access}. 
In assessing the effectiveness of such programs, it would be useful to have a way of characterizing students' interactions and connections within the community. 

Social Network Analysis (SNA) is a quantitative approach used to measure and explore social interactions and communication  between individuals within a group, or network. 
Typically, these networks are created using survey data where students are explicitly asked to name individuals with whom they interact in some manner. We present a novel application of SNA to networks created from qualitative analysis of classroom video data.

In this paper, we briefly present the Social Network Analysis framework and how we have adapted it for use with classroom video data. 
Then, we present a preliminary analysis of our data and discuss how SNA could be applied to these data to address additional research questions.

\section{Social Network Analysis (SNA)}\label{sec:theory} 
Social Network Analysis (SNA) has its roots in quantitative sociology and is built on the idea of structural centrality~\cite{Freeman1979}. 
The core of all network analysis is the identification of \emph{nodes} and \emph{ties}. In a social network, the \emph{nodes} represent individual people in the network.
A \emph{tie} is a link between two nodes and  typically represents an interaction or communication transfer between two people. Sometimes \emph{ties} are directional, where the person who initiates the interaction is called the \emph{source} and the person who receives the information is the \emph{target}. 

In network graphs, nodes are represented as dots and ties are represented as lines between two nodes, or as arrows if directional information is included. For networks where there are multiple ties between nodes, the number or thickness of the ties is used to represent the relative strength of the interaction between nodes.

In recent years, the Physics Education Research (PER) community has taken up SNA as a tool to characterize participation within communities of interest (e.g., a physics learning center~\cite{Brewe2012} or the PER community itself~\cite{Anderson2017}). 
In addition, several studies have used SNA to quantify interactions in order to explore their impact on other constructs relevant to learning (e.g., course grades~\cite{Bruun2013}, self-efficacy~\cite{Dou2016}, persistence~\cite{Zwolak2017}). 

In most of these studies~\cite{Brewe2012,Bruun2013,Dou2016,Zwolak2017}, ties were identified using self-reported data from a survey where participants name individuals with whom they interact.  
We  propose  an application of SNA where ties are instead extracted from qualitative analysis of classroom video data. 

In the following sections, we discuss the context of our study~(\ref{sec:context}), describe how we identified ties from our data~(\ref{sec:data}), present some preliminary analysis~(\ref{sec:analysis}), and discuss additional ways that we plan to apply SNA to these data~(\ref{sec:conclusions}).

\section{What is IMPRESS?}\label{sec:context}
The context for the present study is the Integrating Metacognitive Practices and Research to Ensure Student Success (IMPRESS) summer program, which is a two-week program for matriculating Rochester Institute of Technology (RIT) students who are first generation students and/or deaf/hard of hearing students (DHH)~\cite{Franklin2017IMPRESS}. 
 This program is designed to serve as a bridge program for students to learn how to reflect on, evaluate, and change their own thinking through intensive laboratory experiments, reflective practices, and discussion both in small groups (3-4 students) and with the whole class (20 students).  
The main objectives of the IMPRESS program are to engage students in authentic science practice, to facilitate the development of a supportive community, and to help the students reflect on the science and themselves in order to strengthen their learning habits and lead them to a stronger future in STEM fields. 

We are interested in characterizing and exploring this developing community. 
While it is difficult to track overall community development by simply observing interactions, the quantitative aspect of SNA allows us to better characterize the formation and evolution of the IMPRESS community. 
We analyzed how the patterns of interactions vary over the course of the activity for the different groups. 
In addition, we distinguish between the on-topic and off-topic interactions to help us characterize the amount of time students are  engaged in the science versus time they are building social communities, noting that both of these practices support community building in science classrooms~\cite{Irving2014AdLab}.

\section{Data Collection and Coding}\label{sec:data}
Prior to program, all students were assigned pseudonyms which we used in all subsequent analysis.  
As a part of their application, students self-identified according to gender, ethnicity, and hearing status. 
These data were linked with each pseudonym to allow for analysis along these lines. 
  
Cameras were set up at four (out of five) tables to observe activities and interactions. 
For this preliminary analysis, we focused on a single small group activity on the ninth day of the program where students were trying to come up with equations to represent the carbon cycle based on their work and experiments earlier in the program. 
While the students were working on the assignment, the instructor and two learning assistants (who were previous IMPRESS students) circulated the room interacting with different groups. 

After an inspection of the video data, we chose two-minute clips as our unit of analysis. 
We began our coding as the instructor finished the instructions for the task and ended as he introduced the next task (42 minutes, 21 clips). 
Interactions were coded using Behavioral Observation Research Interactive Software (BORIS) \cite{BORIS}. 

For each two-minute clip, we coded only whether an interaction between two individuals occurred, but not how many times it occurred. 
Moreover, 
we considered all interactions to be directional with an identified source (the person sending information) and targets (individuals that received the information). 
We developed a detailed codebook where we described what counted as an interaction and how it might vary for hearing and DHH individuals.

In addition, we coded each interactions as either being {\it on-topic} or {\it off-topic}. 
{\it On-topic} interactions were those involving conversations related to the activity, to the IMPRESS program or to STEM education; all other interactions not related to education were classified as {\it off-topic}. 

Recognizing the occurrence of interactions and classifying them as on- or off-topic may depend on the interpretation of the codebook. 
To minimize the effect of such interpretations, we conducted inter-rater reliability tests. 
Seven random clips of the video data were coded and compared by three coders with initial agreement of 88\%. 
Discrepancies were discussed and the definitions of the coding schemes were revised.  
The clips were then recoded by all three coders with the new coding definitions with 98\% agreement. 
The remainder of the episode was coded by one of the coders.


\section{Analysis and Discussion}\label{sec:analysis}
For our preliminary analysis, we focused on the interactions within each small group. 
Thus, we removed all ties to instructors or between members of different groups. 
The groups were self-selected and had been together since Day 4 of the program. 
All groups had at least one student who identified as DHH. 
Groups 1 and 4 were mixed gender; Group 2 was all male; and Group 3 was all female. 
Groups 1, 2, and 4 each had four people, whereas Group 3 only had three.

In our analysis, we considered two cases: (1) interactions within individual two-minute clips (as described in Sec.~\ref{sec:data}) and (2) weighted interactions over multiple clips (averaged data). 
For the weighted network, we aggregated four (non-overlapping) clips per bin (8 minutes). Due to an odd number of clips, the last bin contained five clips (10 minutes). 
We then plotted these individual and averaged data over time.

\begin{figure}
\includegraphics[width=0.9\linewidth]{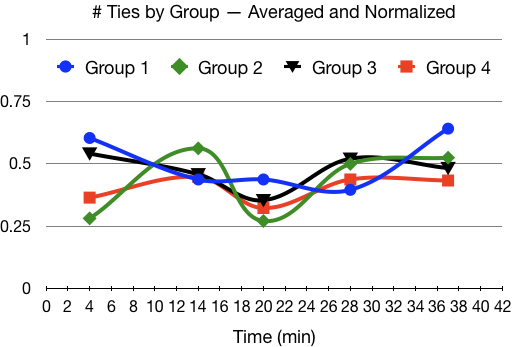}
\caption{Number of on-topic ties by group normalized and averaged by the total possible ties in each bin. The first four bins contain 4 clips each (8 min) and the last bin contains 5 clips (10 min). \label{fig:all-normalized}}
\end{figure}

The first point of comparison between groups was how many interactions occurred overall. Given the differing numbers of group members, we normalized by the total possible interactions for each group.
For a two-minute clip, there are 24 possible interactions for a 4-person group and 12 possible interactions for a 3-person group. The data for the full episode are shown in Table~\ref{tab:counts}.  
Based on this table, Groups 2 and 4 appear to interact less than Groups 1 and 3, but examining averaged and normalized data over time (see Fig.~\ref{fig:all-normalized}) shows no clear discernible patterns of difference between groups.

\begin{table}[b]
\caption{Total number of ties during the episode by group.\label{tab:counts}}
\begin{ruledtabular}
\begin{tabular}{lcccc}
 & Group 1 & Group 2 & Group 3 & Group 4\\ 
 \hline
Total \# of Ties &257	&218	&119	&203\\
\# of Possible Ties&504	&504	&252	&504\\
\# of Ties (normalized)&0.51	&0.43	&0.47	&0.40
\end{tabular}
\end{ruledtabular}
\end{table}

\begin{figure}[t]
\includegraphics[width=0.9\linewidth]{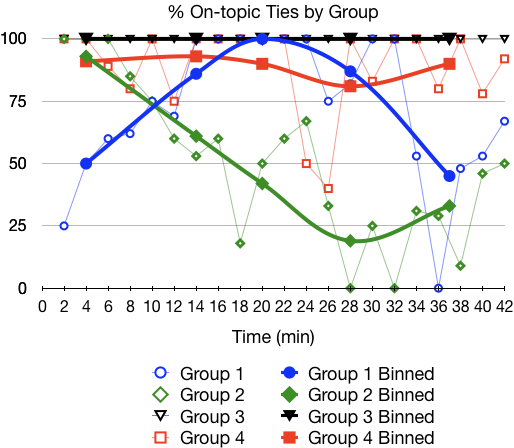}
\caption{The unfilled circles connected by thin lines represent the percentage of on-topic ties in each individual clip by group. The solid dots connected by thick lines represent the average percentage of on-topic ties in each bin by group. The first four bins contain 4 clips each (8 min) and the last bin contains 5 clips (10 min). \label{fig:on-combined}}
\end{figure}

We also wanted to explore how the frequency of {\it on-topic} and {\it off-topic} interactions changed over time. 
Figure~\ref{fig:on-combined} shows the percentage of {\it on-topic} ties for each group over time, both for individual clips and averaged data.  
Unlike the overall number of ties, we see clearly different behavior between groups when looking at the percentage of {\it on-topic} ties.

Given that we began as the instructor was finishing the instructions for the task, we expected to see a relatively high (or increasing) number of {\it on-topic} interaction as the students began working on the activity and then an increase in {\it off-topic} interactions as groups began to wrap-up the activity. 
The behavior of Groups 1 and 4 mostly follow this expected pattern.
However, Groups 2 and 3 exhibit more unexpected behavior. 
The trend for Group 2 is strongly declining towards {\it off-topic} most of the time while Group 3 stays {\it on-topic} the whole time. 

\begin{figure*}
\begin{minipage}{0.3\linewidth}
\begin{minipage}{0.45\linewidth}
Group 1\\
\includegraphics[width=\linewidth]{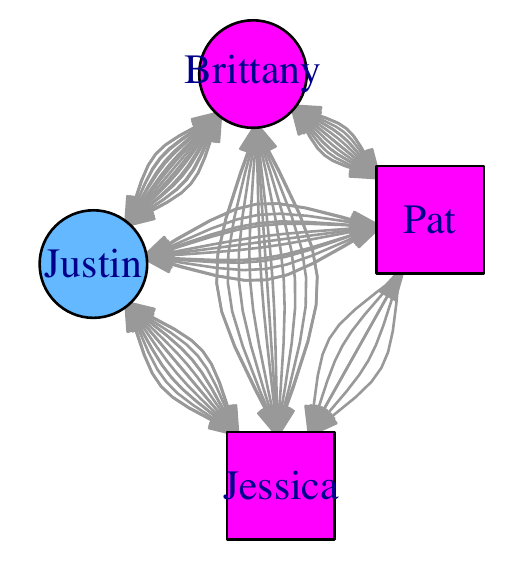}
\end{minipage}
\begin{minipage}{0.45\linewidth}
Group 3\\
\includegraphics[width=\linewidth]{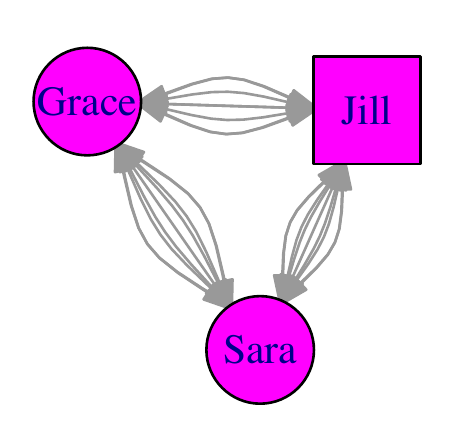}
\end{minipage}
\begin{minipage}{0.45\linewidth}
Group 2\\
\includegraphics[width=\linewidth]{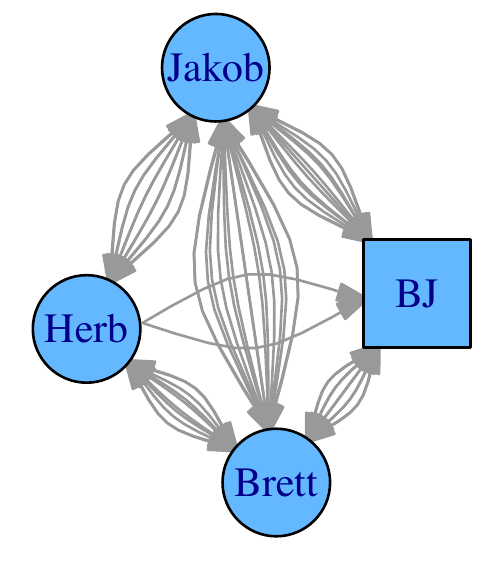}
\end{minipage}
\begin{minipage}{0.45\linewidth}
Group 4\\
\includegraphics[width=\linewidth]{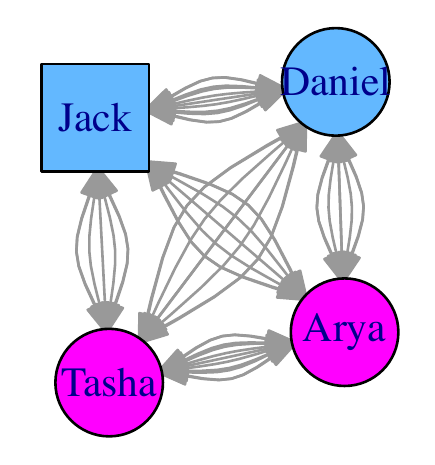}
\end{minipage}\\
(a) all ties
\end{minipage}
\hfill
\vline
\hfill
\begin{minipage}{0.3\linewidth}
\begin{minipage}{0.45\linewidth}
Group 1\\
\includegraphics[width=\linewidth]{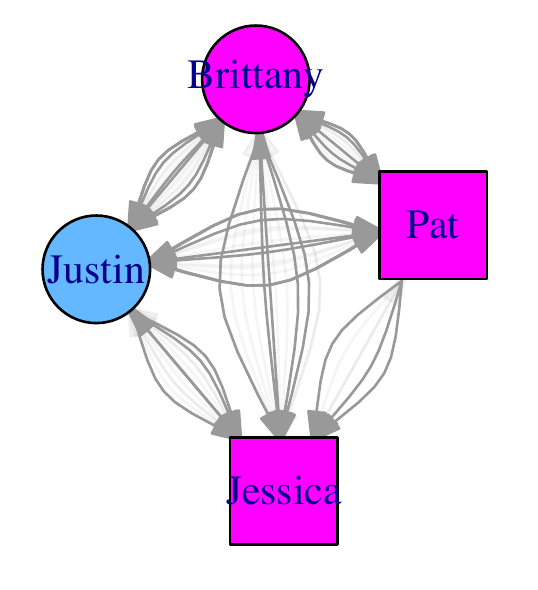}
\end{minipage}
\begin{minipage}{0.45\linewidth}
Group 3\\
\includegraphics[width=\linewidth]{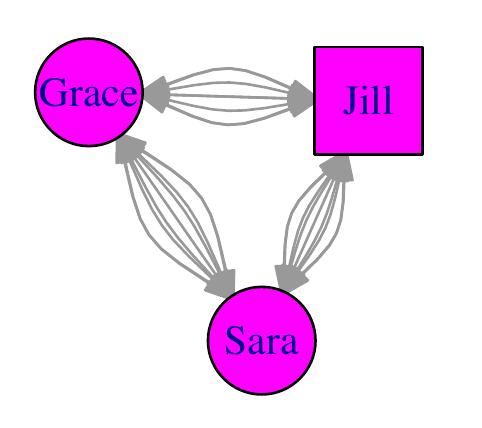}
\end{minipage}
\begin{minipage}{0.45\linewidth}
Group 2\\
\includegraphics[width=\linewidth]{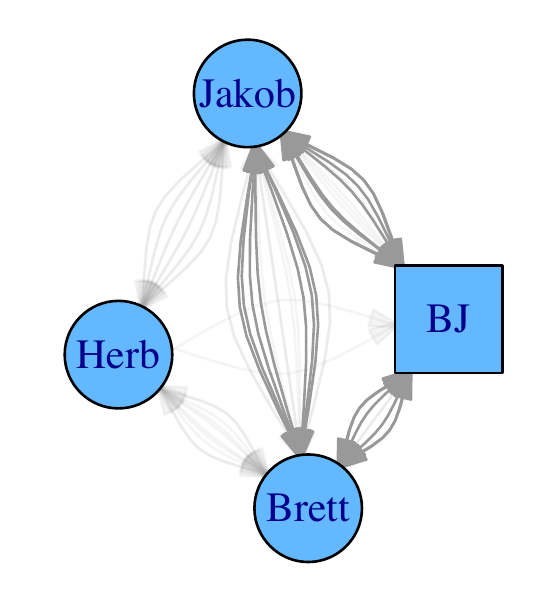}
\end{minipage}
\begin{minipage}{0.45\linewidth}
Group 4\\
\includegraphics[width=\linewidth]{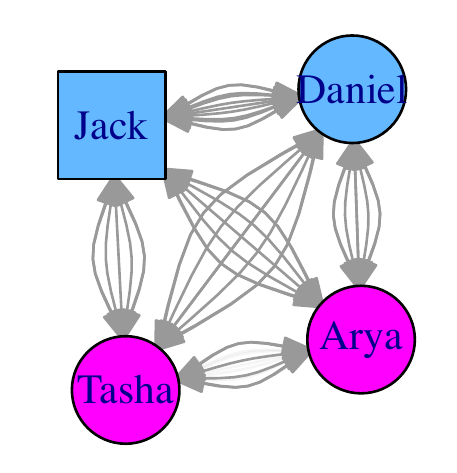}
\end{minipage}\\
(b) on-topic ties
\end{minipage}
\hfill
\vline
\hfill
\begin{minipage}{0.3\linewidth}
\begin{minipage}{0.45\linewidth}
Group 1\\
\includegraphics[width=\linewidth]{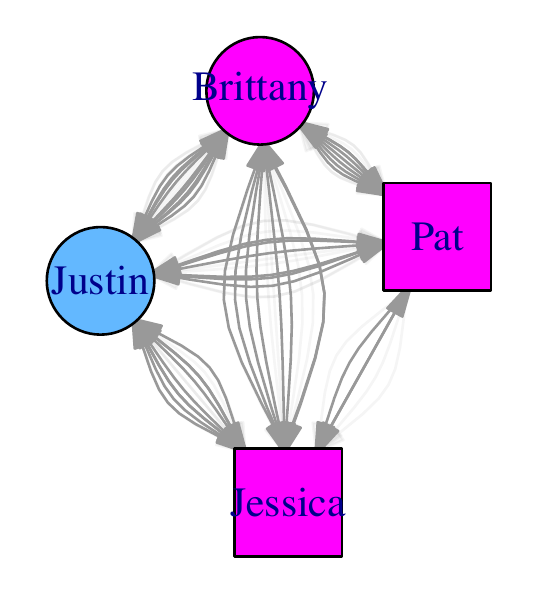}
\end{minipage}
\begin{minipage}{0.45\linewidth}
Group 3\\
\includegraphics[width=\linewidth]{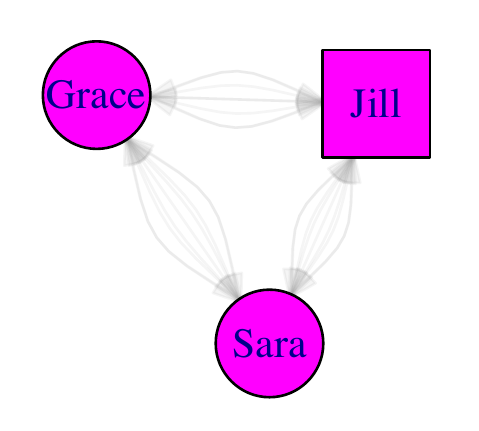}
\end{minipage}
\begin{minipage}{0.45\linewidth}
Group 2\\
\includegraphics[width=\linewidth]{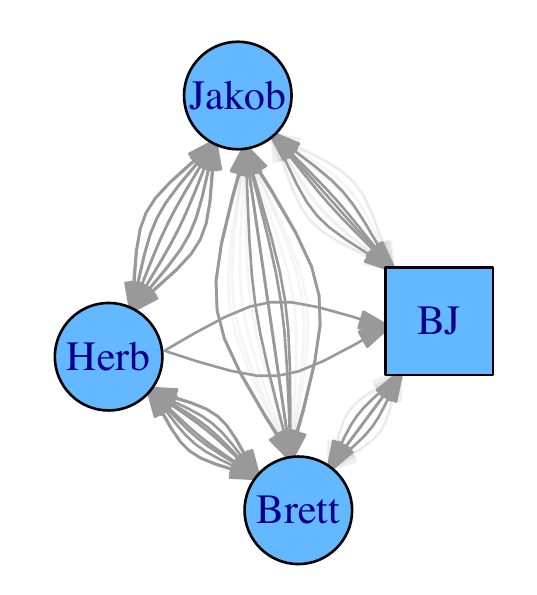}
\end{minipage}
\begin{minipage}{0.45\linewidth}
Group 4\\
\includegraphics[width=\linewidth]{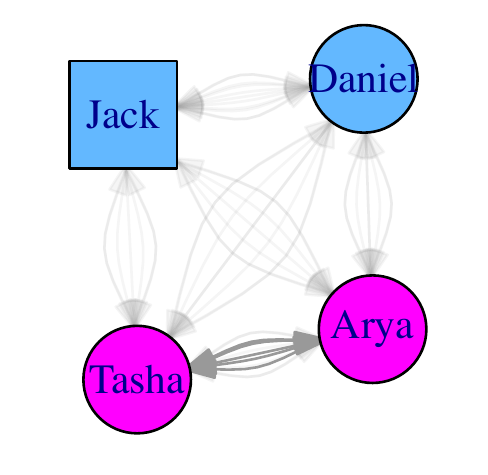}
\end{minipage}\\
(c) off-topic ties
\end{minipage}
\caption{Network graph for the last bin (32-42 min) (a) showing all ties, (b) highlighting on-topic ties, and (c) highlighting off-topic ties. Node color denotes the gender of the student (blue = male; pink = female). Node shape indicates hearing status (circle = hearing; square = DHH). \label{fig:network}}
\end{figure*}

To get a better sense for how individuals might be contributing to the overall patterns, we created weighted network maps for each bin, where each arrow represented a directed interaction and gender and hearing status were included as color and shape, respectively. 
Figure~\ref{fig:network} shows the network graphs for the last bin (32-42 minutes) with (a) all ties, (b) highlighting on-topic ties, and (c) highlighting off-topic ties. 
This bin was chosen to highlight the differences in the groups, though we find it to be a good representative of the groups' behavior at other times as well. 

While it appears in Fig.~\ref{fig:network}(a) that the students interacted with all group members overall, Figs.~\ref{fig:network}(b) and (c) indicate that some students are primarily interacting  only {\it on-topic} (e.g., Jack and Daniel) or only {\it off-topic} (e.g., Herb).

When looking at the {\it on-topic} ties in Fig.~\ref{fig:network}(c), the most striking thing is that Herb, in Group 2, does not initiate or receive any {\it on-topic} interactions during this ten-minute time period. 
This suggests Herb's involvement in the group's dynamic may be one of the reasons for Group 2's {\it off-topic} trend towards {\it off-topic} conversations that was noted in Fig.~\ref{fig:on-combined}.

\begin{figure}[t]
\includegraphics[width=0.9\linewidth]{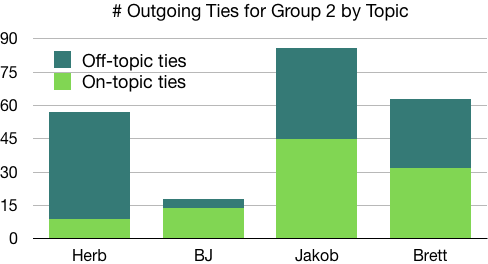}
\caption{Number of outgoing ties for each member of Group 2 based on whether they were on-topic or off-topic. \label{fig:group2-outgoing}}
\end{figure}

In order to more closely examine Herb's role in the group, we  compared the number of outgoing ties for {\it on-topic} and {\it off-topic} interactions for each student in Group 2. 
As can be seen in  Fig.~\ref{fig:group2-outgoing}, Herb has more outgoing {\it off-topic} ties~(48) than any other group member, including Jakob~(41) and Brett~(31). 
Even more telling is that he initiates significantly more off-topic~(84\%) interactions than on-topic interactions~(16\%).  
This provides additional evidence that Herb may be at least part of the cause of Group 2's {\it off-topic} trend.

Figure~\ref{fig:group2-outgoing} also shows that although BJ initiated more {\it on-topic} than {\it off-topic} interactions, he contributed to the discussion much less overall than the other group members.
Since BJ was the only member of Group~2 who identified as DHH, this result suggests a closer look at 
how DHH students participate differently than hearing students.

In Fig.~\ref{fig:network}(c), we see the students in Group~3 exclusively interacting {\it on-topic}. 
In fact, this group had no {\it off-topic} interactions amongst themselves during the entire activity (see~Fig.~\ref{fig:on-combined}), which suggests that this table may not be actively developing  as a community. 

We plan to perform similar analyses for other activities to determine if the behaviors seen here are typical of these groups. 
We also plan to compare individuals' behavior to earlier in the program when they were in different groups. 

\section{Conclusions and Future Work}\label{sec:conclusions}
Social Network Analysis is wonderful tool to quantify interactions between individuals in a group.  
While SNA has historically used data from surveys where participants self-report connections, it has not been used for analysis of video data. 
We applied SNA to one activity from the IMPRESS summer program at RIT to test our approach and begin to explore whether the goals of the program were being met. 

Using SNA with classroom video data allowed us to visualize and analyze the interactions between groups and individuals within our network. 
More importantly, our distinction between {\it on-topic} versus {\it off-topic} interactions helped us to begin to assess whether the group behavior was as expected - with a balance of engagement with science and community formation - or whether there were some groups or individuals that exhibited unexpected behavior.    
Graphing {\it on-topic} ties over time identified one table that may not be  developing as a community in the way the program intended. 
The network graphs pointed to a possible cause of another group's trend toward {\it off-topic} interactions: Herb. 
Looking at directional information supported this conclusion but also revealed a more complex group dynamic.

We have shown that it is possible to extract data from classroom video to create and analyze social networks. 
Our preliminary analysis only begins to scratch the surface of what Social Network Analysis can do. 
As mentioned earlier, we intend to extend this analysis to other activities at various points in the program. 
We will use these data and SNA to explore how individual and group interactions change over time. 
We will also look more closely at how communication patterns are impacted by different aspects of student identity such as gender, ethnicity, hearing status, or being a first-generation college student. 
Finally, we plan to extend this analysis to look beyond the interactions within individual groups to their interactions with instructors and other groups.

Thus, Social Network Analysis will allow us to characterize the participation of students in the IMPRESS community to better assess how this program is achieving its goal of  helping students to develop as STEM professionals.\\

\acknowledgments{This project has been funded by NSF PHY 1344247, the DePaul University Graduate Research Fund (GRF), and a PERLOC travel grant.}

\bibliographystyle{apsrev}  	
\bibliography{NetworkBibliography}  	

\end{document}